\documentclass[preprint]{aastex}

 \usepackage{emulateapj5}
 \usepackage{apjfonts}

\newcommand{\Dnu}{\mbox{$\Delta \nu$}}
\newcommand{\acena}{\mbox{$\alpha$~Cen~A}}
\newcommand{\acenb}{\mbox{$\alpha$~Cen~B}}

\newcommand{\afor}{\mbox{$\alpha$~For}}
\newcommand{\bhyi}{\mbox{$\beta$~Hyi}}
\newcommand{\gser}{\mbox{$\gamma$~Ser}}
\newcommand{\baql}{\mbox{$\beta$~Aql}}
\newcommand{\dpav}{\mbox{$\delta$~Pav}}

\newcommand{\nuind}{\mbox{$\nu$~Ind}}

\newcommand{\cms}{\mbox{cm\,s$^{-1}$}}

\newcommand{\muHz}{\mbox{$\mu$Hz}}

\slugcomment{To appear in ApJ}

\shorttitle{Amplitudes of solar oscillations}
\shortauthors{Kjeldsen et al.}

\begin{document}

\title{The amplitude of solar oscillations using stellar techniques}

\author{
Hans~Kjeldsen,\altaffilmark{1}
Timothy R. Bedding,\altaffilmark{2}
Torben Arentoft,\altaffilmark{1}
R.~Paul~Butler,\altaffilmark{3}
Thomas~H.~Dall,\altaffilmark{4}
Christoffer~Karoff,\altaffilmark{1}
L\'aszl\'o L. Kiss,\altaffilmark{2}
C.~G.~Tinney\altaffilmark{5} and
William~J.~Chaplin\altaffilmark{6}
}

\altaffiltext{1}{Danish AsteroSeismology Centre (DASC), Department of
Physics and Astronomy, University of Aarhus, DK-8000 Aarhus C, Denmark;
hans@phys.au.dk, toar@phys.au.dk, karoff@phys.au.dk}

\altaffiltext{2}{School of Physics A28, University of Sydney, NSW 2006,
Australia; bedding@physics.usyd.edu.au, laszlo@physics.usyd.edu.au}

\altaffiltext{3}{Carnegie Institution of Washington,
Department of Terrestrial Magnetism, 5241 Broad Branch Road NW, Washington,
DC 20015-1305; paul@dtm.ciw.edu}

\altaffiltext{4}{Gemini Observatory, 670 N. Aohoku Pl., Hilo, HI 96720,
USA; tdall@gemini.edu}

\altaffiltext{5}{School of Physics, University of NSW, 2052, Australia;
cgt@phys.unsw.edu.au}

\altaffiltext{6}{School of Physics and Astronomy, University of Birmingham,
Edgbaston, Birmingham B15 2TT, UK; wjc@bison.ph.bham.ac.uk}

\begin{abstract} 
The amplitudes of solar-like oscillations depend on the excitation and
damping, both of which are controlled by convection.  Comparing
observations with theory should therefore improve our understanding of the
underlying physics.  However, theoretical models invariably compute
oscillation amplitudes relative to the Sun, and it is therefore vital to
have a good calibration of the solar amplitude using stellar techniques.
We have used daytime spectra of the Sun, obtained with HARPS and UCLES, to
measure the solar oscillations and made a detailed comparison with
observations using the BiSON helioseismology instrument.  We find that the
mean solar amplitude measured using stellar techniques, averaged over one
full solar cycle, is $18.7 \pm 0.7$\,\cms{} for the strongest radial modes
($l=0$) and $25.2\pm 0.9$\,\cms{} for~$l=1$.  In addition, we use
simulations to establish an equation that estimates the uncertainty of
amplitude measurements that are made of other stars, given that the mode
lifetime is known.  Finally, we also give amplitudes of solar-like
oscillations for three stars that we measured from a series of short
observations with HARPS (\gser, \baql{} and \afor), together with revised
amplitudes for five other stars for which we have previously published
results (\acena, \acenb, \bhyi, \nuind{} and \dpav).
\end{abstract}


\section{Introduction}

The list of stars in which solar-like oscillations have been observed is
growing rapidly, thanks to improvements in high-precision velocity
measurements (for a recent review see \citealt{B+K2007b}).  While most
excitement centres on the frequencies of the oscillations, there is also
considerable interest in the amplitudes
\citep[e.g.,][]{ChD+F83,K+B95,HBChD99,H+G2002,Hou2006,SGA2005,SBG2007,SGT2007,SBG2008}.
This is because the excitation and damping are both controlled by
convection and so the study of oscillation amplitudes will hopefully lead
to an improvement in our understanding of the underlying physics.

Theoretical calculations of oscillation amplitudes, such as those cited
above, are calibrated with reference to the Sun.  It is therefore crucial
to establish the amplitudes of the solar oscillations.  There is an
abundance of solar observations from helioseismology projects such as BiSON
(Birmingham Solar Oscillations Network) and GOLF (Global Oscillations at
Low Frequencies).  However, these velocity measurements are made using a
single spectral line, such as sodium ($\lambda5896$\AA, used by GOLF) or
potassium ($\lambda7699$\AA, BiSON), and the measured oscillation amplitude
depends on the height in the solar atmosphere at which the spectral line is
formed (e.g., \citealt{IMP89,BSG2005,Hou2006}).  Velocity measurements of
other stars, on the other hand, employ a wide wavelength range that
includes many spectral lines, mostly from neutral iron.  The iron lines are
formed lower in the atmosphere than sodium and potassium and so we might
expect solar amplitudes measured using stellar techniques to be lower than
those from helioseismic measurements such as BiSON and GOLF\@.  On the
other hand, BiSON and GOLF measure velocities using a resonance scattering
cell, whereas the stellar technique involves measuring line centroids.
These approaches have different sensitivities as a function of the mode
angular degree~$l$ \citep{ChD89}, which must be corrected for when
interpreting measured amplitudes.

We have recently pointed out the importance of establishing the oscillation
amplitude of the Sun using stellar techniques \citep{KBB2005}.  In that
paper, we reported a limited set of observations of the solar oscillations
obtained from spectra of the full moon.  The results gave some support to
the conclusion that Fe~I measurements give lower amplitudes than both GOLF
and BiSON, but we emphasized the desirability of obtaining more
measurements of the Sun, in order to better calibrate the relationship
between stellar and solar amplitudes.  That is the main purpose of this
paper.

\section{Observations}

We have obtained solar spectra by observing the daytime sky.  Although
detailed line profiles of daytime spectra show differences from the actual
solar spectrum \citep{GTB2000}, and the bisector varies with solar angle
\citep{DSA2006}, there is no reason to suspect any systematic effects on
mean velocities on the timescale of the 5-minute oscillations.

\begin{figure*}
\epsscale{0.5}
\plotone{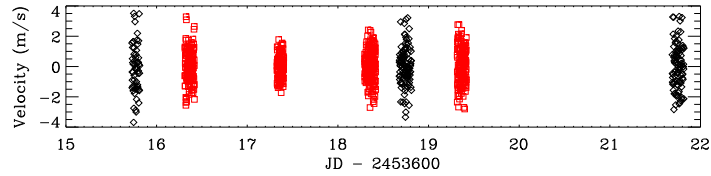}
\caption[]{\label{fig.series} Velocity time series of the Sun measured from
  the blue sky by HARPS (red squares) and UCLES (black diamonds).  }
\end{figure*}

We made the observations in 2005 September during a dual-site
asteroseismology campaign on the star \bhyi{} \citep{BKA2007}.
Figure~\ref{fig.series} shows the full set of data.
At the European Southern Observatory on La Silla in Chile we used the HARPS
spectrograph (High Accuracy Radial velocity Planet Searcher) with the 3.6-m
telescope.\footnote{Based on observations collected at the European
Southern Observatory, La Silla, Chile (ESO Programme 075.D-0760)}\@.  A
thorium emission lamp was used to provide a stable wavelength reference.
We obtained 626 spectra over 9.7\,hr spread over four consecutive
afternoons, with typical exposure times of 25\,s and a median cadence of
one exposure every 56\,s (which corresponds to a Nyquist frequency of
8.9\,mHz).  The velocities were processed using the standard HARPS
pipeline.

At Siding Spring Observatory in Australia we used UCLES (University College
London Echelle Spectrograph) with the 3.9-m Anglo-Australian Telescope
(AAT).  An iodine absorption cell was used to provide a stable wavelength
reference, with the same setup that we have previously used with this
spectrograph \citep{BBK2004}.  With UCLES we obtained 265 spectra of the
Sun over about 8.5\,hr, spread over three non-consecutive afternoons.
Typical exposure times were 45--60\,s, with a median cadence of one
exposure every 115\,s (which corresponds to a Nyquist frequency of
4.35\,mHz).

In addition, we have analysed velocities obtained by BiSON (Birmingham
Solar Oscillations Network) over the first 10 days of 2005 September, which
covers the period in which the UCLES and HARPS data were taken.  We also
examined a 14-year time series (1992.6 to 2006.9) from the BiSON archive,
which allows us to follow variations in amplitude with the solar activity
cycle (see \S\ref{sec.bison14}).  The BiSON data have 40-second sampling.
They had also been high-pass filtered by subtracting a moving mean
calculated in a boxcar of length 1000\,s.  We have corrected for this by
dividing the power spectra by the transfer function of this filter.

\subsection{Observations of \gser, \baql{} and \afor} \label{sec.stars.obs}

As mentioned above, the night-time target for this asteroseismology
campaign was the star \bhyi.  However, \bhyi{} was not accessible to HARPS
for the first part of each night due to hour-angle restrictions on the ESO
3.6-m telescope.  We used this time to observe two other stars, taken from
the list of targets presented by \citet{BKR96}.  The first was \gser{}
(HR~5933, HD~142860, HIP~78072), which has spectral type F6\,IV-V and
magnitude $V=3.85$, and was observed for about 75 minutes at the start of
each of three consecutive nights.  The second star was \baql{} (HR~7602,
HD~188512, HIP~98036), spectral type G8\,IV and magnitude $V=3.71$, which
was observed on four consecutive nights for periods of 65, 95, 10 and 70
minutes, respectively.

This limited set of observations was not sufficient to allow us to measure
oscillation frequencies, but we were able to estimate the amplitudes, in
the same way as we have previously done for the star \dpav{}
\citep{KBB2005}.  In addition, we also analysed observations of the star
\afor{} (HR~963, HD~20010, HIP~14879), which has spectral type F8\,IV and
magnitude $V=3.85$.  Those observations were taken with HARPS in 2007
January for about 20-30 minutes at the start of each of seven consecutive
nights, during a multi-site campaign on Procyon (T. Arentoft et al., in
prep.).  Note that the behaviour of the line bisectors for these
observations of \gser{} and \baql{} were already discussed by
\citet{DSA2006}.

\section{Analysis and results}

\subsection{Time series and power spectra}

\begin{figure*}
\epsscale{0.8}
\plotone{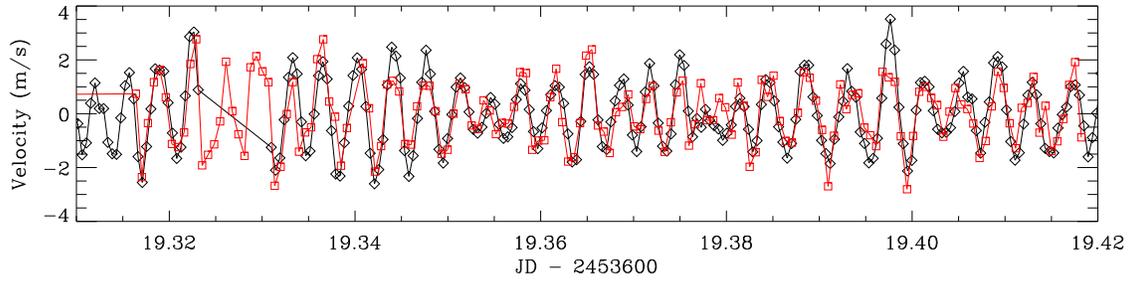}
\caption[]{\label{fig.series.closeup} Close-up showing 160\,min of the
HARPS (red squares) together with contemporaneous measurements by BiSON
(black diamonds).  }
\end{figure*}

The first step in the analysis of the solar velocities from HARPS and UCLES
was to remove slow trends from each data set by a high-pass filter that
only affected frequencies below 1\,mHz.  Figure~\ref{fig.series.closeup}
shows a close-up comparison of HARPS and BiSON for one afternoon of data
(160\,min).  There is excellent agreement and we see that the two data sets
are of comparable quality.  The UCLES data have somewhat poorer precision
per data point.

\begin{figure*}
\epsscale{0.5}
\plotone{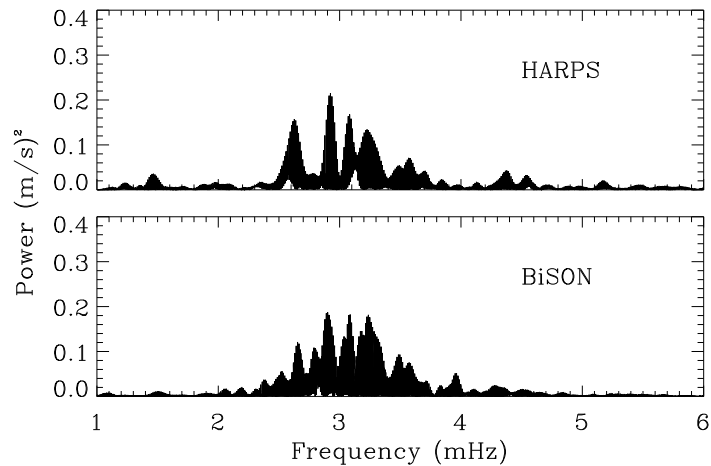}
\caption[]{\label{fig.power.harps} Power spectrum of the HARPS velocities
  of the Sun (upper), and of the BiSON series taken at the same times
  (lower).  }
\end{figure*}

\begin{figure*}
\epsscale{0.5}
\plotone{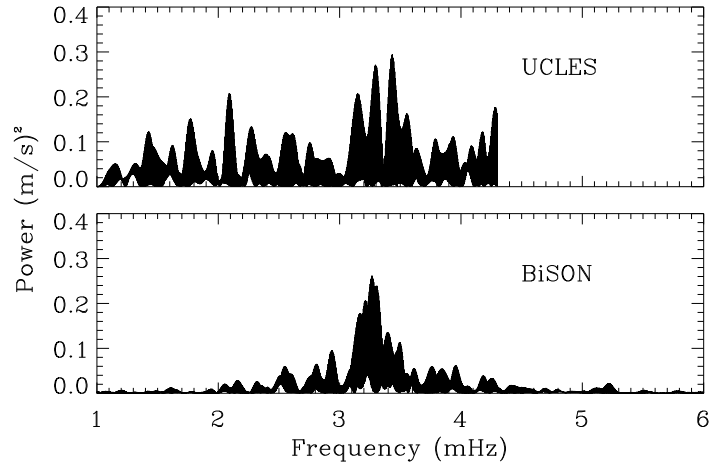}
\caption[]{\label{fig.power.ucles} Same as Fig.~\ref{fig.power.harps}, but
  for UCLES and the corresponding BiSON data.  }
\end{figure*}

The next step was to calculate the power spectra of the velocities.  For
HARPS and UCLES, we used the measurement uncertainties, $\sigma_i$, as
weights in calculating the power spectrum (according to $w_i =
1/\sigma_i^2$).  No uncertainty estimates were available for the BiSON
velocities and so all points were given equal weight.  The power spectrum
of the HARPS time series is shown in the upper panel of
Fig.~\ref{fig.power.harps}, while the lower panel shows the spectrum of the
segment of BiSON data taken at the same times.  We see that the HARPS and
BiSON power spectra are very similar.  Figure~\ref{fig.power.ucles} shows a
similar analysis for the UCLES data and the corresponding BiSON segment.
The UCLES power spectrum has higher noise but, taking this into account, we
again see good agreement between the two.  A comparison of
Figures~\ref{fig.power.harps} and~\ref{fig.power.ucles} show that the solar
power spectrum was quite different for the two time series, reflecting the
intrinsic variations in the solar oscillations.  Given that the HARPS data
have lower noise than the UCLES data, we do not show results for UCLES in
the remainder of the analysis.


\subsection{Calculating smoothed amplitude spectra}  \label{sec.amp}

The amplitudes of individual modes are affected by the stochastic nature of
the excitation and damping.  To measure the oscillation amplitude in a way
that is independent of these effects, we have followed the method
introduced by \citet{KBB2005}.  In brief, this involves the following
steps: (i)~heavily smoothing the power spectrum (by an amount defined in
the next paragraph), to produce a single hump of excess power that is
insensitive to the fact that the oscillation spectrum has discrete peaks;
(ii)~converting to power density by multiplying by the effective length of
the observing run (which we calculate as the reciprocal of the area under
the spectral window in power); (iii)~fitting and subtracting the background
noise; and (iv)~multiplying by $\Delta\nu/c$ and taking the square root, in
order to convert to amplitude per oscillation mode.  Here, $\Delta\nu$ is
the large frequency separation, which has a value of 135\,\muHz{} in the
Sun, and~$c$ is a factor that measures the effective number of modes per
order.

The amount of smoothing affects the exact height of the smoothed amplitude
spectrum.  To establish a standard that allows comparisons, we propose that
smoothing be done by convolving (in power) with a Gaussian having a full
width at half maximum of $4\Delta\nu$.  We have adopted this convention for
all the figures and measurements in this paper.

\subsubsection{The value of $c$}

In our previous work, we used a value of $c=3.0$
\citep{KBB2005,BBC2006,BKA2007}.  However, $c$ depends on the sensitivity
of the observations to the various low-degree modes, which in turn depends
on the method used to observe the oscillations.  We therefore reconsider
the calculation of~$c$, as follows.  Let $S_l/S_0$ be the spatial response
of the observations to modes with degree $l$, relative to those with $l=0$.
Table~4 of \citet{ChD89} allows us to calculate $S_l/S_0$ for BiSON
velocity observations and these are shown in Table~\ref{tab.spatial}.  We
have also calculated values for velocity observations using the stellar
technique, using the results in \citet{BKR96} for an adopted mean
wavelength of 550\,nm.  These are also shown in Table~\ref{tab.spatial}.
The difference from the BiSON values arises from the way the BiSON
resonance scattering cell measures velocities, as discussed in detail by
\citet{ChD89}.  Finally, Table~\ref{tab.spatial} also shows the response
factors for intensity measurements at the three wavelengths of the SoHO
VIRGO instrument, again derived from the
results in \citet{BKR96}.\footnote{The values in Table~\ref{tab.spatial}
are summed over all $m$ values.  To calculate response functions for
individual $m$ values, as would be observed in rotationally split modes,
see \citet{G+S2003} and references within.}

The last row of Table~\ref{tab.spatial} shows the value of $c$, calculated
as
\begin{equation}
  c = \sum_{l=0}^4 \left(S_l/S_0\right)^2.
\end{equation}
These are the values that we will adopt in future.  Note that the value of
$c=3.0$ that we used previously was based on normalizing to the mean of
$l=0$ and 1, rather than to $l=0$, as we now propose.  The $l=0$ modes make
a sensible reference because they are not split by rotation, but it is
important to remember that the $l=1$ modes are actually stronger than the
radial modes.

Note that comparing observed amplitudes with theoretical calculations
requires knowledge of the actual value of the spatial response function of
the observations.  For intensity measurements, $S_0=1$ by definition.  For
BiSON, \citet{ChD89} has calculated $S_0 = 0.724$, while for stellar
velocity measurements we have used the results in \citet{BKR96} to derive
$S_0 = 0.712$ (again, for an adopted mean wavelength of 550\,nm).

\subsection{The solar amplitude}  \label{sec.bison14}

\begin{figure*}
\epsscale{0.5}
\plotone{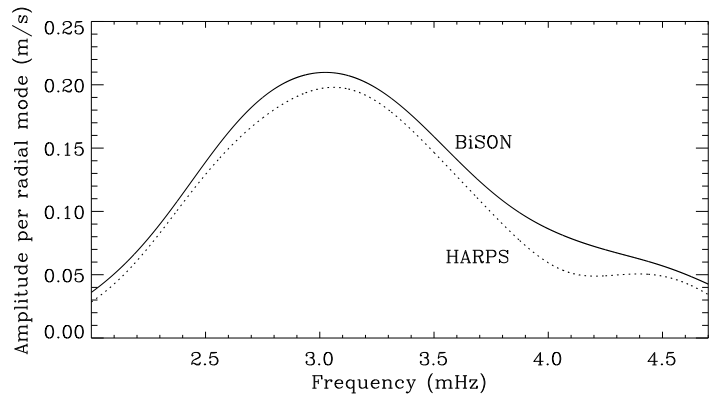}
\caption[]{\label{fig.amp.bison.harps} Amplitude per mode of oscillations in
  the Sun, as measured from the power spectra in
  Fig.~\ref{fig.power.harps}. }
\end{figure*}

The smoothed amplitude spectrum for our solar observations with HARPS is
shown in Fig.~\ref{fig.amp.bison.harps}, together with that from BiSON from
exactly the same observation times.  Note that here we did not use the
weights when calculating the HARPS power spectrum, since these were not
available for BiSON and we wanted to make the window functions exactly the
same.  The amplitude of solar oscillations measured using the stellar
technique is slightly lower than that measured by BiSON, by a factor of
$1.07\pm0.04$.  The error bar comes from the uncertainty in fitting the
background to the HARPS spectrum.  However, we must keep in mind that our
observations of the Sun were made over only a few hours, and at a single
epoch during the 11-year solar cycle.  To correct for this and hence
determine the mean amplitude of the Sun, we have made use of the full set
of BiSON observations.

\begin{figure*}
\epsscale{0.5}
\plotone{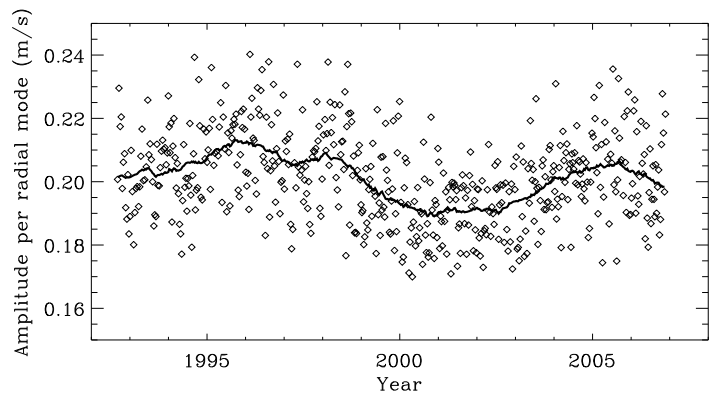}
\caption[]{\label{fig.bison.amp.series} Amplitude of solar oscillations
  measured from BiSON data.  The points were measured from independent
  10-day segments of data and the thick line is after smoothing with a
  boxcar running mean of 50 points (500 days). }
\end{figure*}

We have analysed the 14-year time series from the BiSON archive, measuring
amplitudes in ten-day segments in the way described above.  In practice we
chose all segments to have the same number of data points, which meant that
the lengths of most segments were in the range 8--12\,d, reflecting
variations in the filling factor of the network.  The result is shown in
Fig.~\ref{fig.bison.amp.series}, where each point represents the peak
amplitude in one segment.  We see clearly the variations in amplitude with
the solar activity cycle, as has been well studied from these BiSON data by
\citet{CEI2000,CEI2003}.

Note that the scatter in the amplitudes in Fig.~\ref{fig.bison.amp.series}
is not measurement error, but rather reflects the intrinsic variability of
the modes.  We see that both the amplitude and the scatter on the amplitude
are smallest at solar maximum (in 2001), which is explained by the shorter
mode lifetime during periods of high solar activity.  Any changes in the
excitation rate during the solar cycle would also contribute to amplitude
variability, but no evidence for such changes has so far been reported.

The mean BiSON amplitude over one 11-year solar cycle, measured from the
data in Fig.~\ref{fig.bison.amp.series}, is $20.0\pm0.1$\,\cms.  Combining
this with the result in the previous section, we conclude that the mean
amplitude of the radial oscillations in the Sun, measured using stellar
techniques, is $18.7 \pm 0.7$\,\cms.  This is the quantity that we were
seeking to measure.  It is important to remember that this is based on the
process we used (including the $4\Dnu$-smoothing described in
\S\ref{sec.amp}) and is normalized to the modes with $l=0$.  Values for
non-radial modes can be calculated from the ratios in column~3 of
Table~\ref{tab.spatial}.  For example, the mean peak amplitude for $l=1$
modes is significantly higher than for $l=0$, and both are given in the
first line of Table~\ref{tab.amps}.

\begin{figure*}
\epsscale{0.5}
\plotone{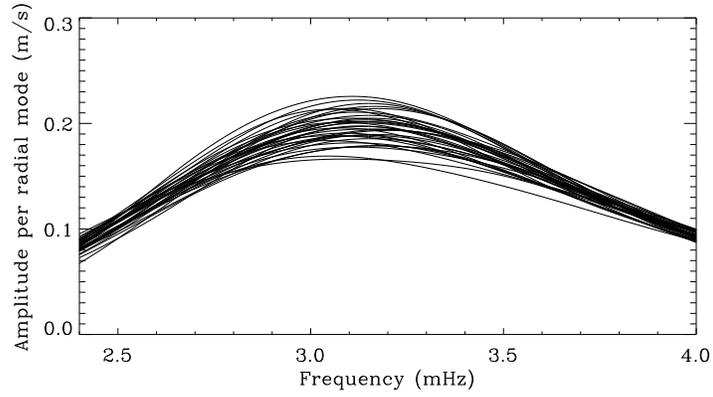}
\caption[]{\label{fig.bison.amp.smoothed} Smoothed amplitude spectra of
  solar oscillations measured from BiSON data in the year 2005.  Each curve
  was measured from an independent 10-day segment of data. }
\end{figure*}

Figure~\ref{fig.bison.amp.smoothed} shows smoothed amplitude spectra for
10-day segments of BiSON data for one full year, starting at 2005.0\@.
This period covers the solar minimum and so the variations are dominated by
the stochastic nature of the excitation and damping, rather than by
variations due to the solar cycle.  Note that the curves show less spread
at the higher frequencies than at the lower frequencies, reflecting the
fact that the mode lifetimes in the Sun increase with frequency.

The mean of these curves is shown in Fig.~\ref{fig.amp.stars} after scaling
so that the peak is $18.7$\,\cms, which is the solar mean as measured with
stellar techniques.

\subsection{Amplitudes for other stars} \label{sec.stars}

\begin{figure*}
\epsscale{0.5}
\plotone{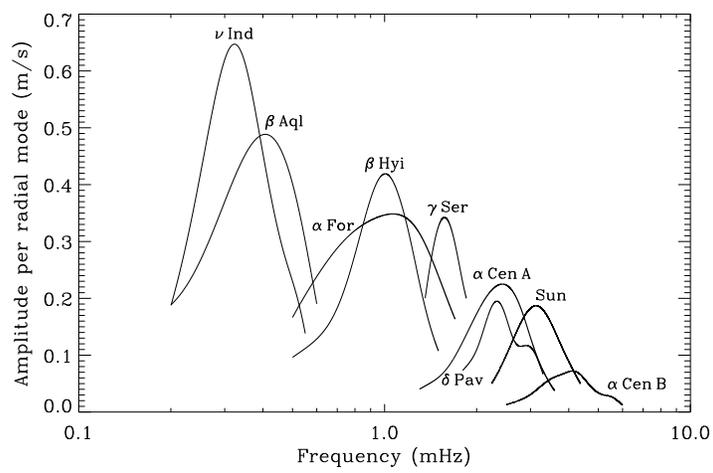}
\caption[]{\label{fig.amp.stars} Smoothed amplitude curves for oscillations
  in the Sun and other stars (see text for details).  }
\end{figure*}

In Figure~\ref{fig.amp.stars} we show the amplitude curves for the three
stars described in \S\ref{sec.stars.obs} (\baql, \afor{} and \gser),
together with five other stars for which we have previously published
amplitudes: \acena, \acenb{} and \dpav{} \citep{KBB2005}, \nuind{}
\citep{BBC2006} and \bhyi{} \citep{BKA2007}.  All the curves were
calculated using the revised method described in \S\ref{sec.amp}.

Table~\ref{tab.amps} summarizes our results for the Sun and these eight
stars.  Column~2 gives the large frequency separation that we used in the
calculation (there are no measurements of \Dnu{} for \gser, \afor, \baql{}
and \dpav{} and so we estimated a value from the stellar parameters).
Column~3 of Table~\ref{tab.amps} gives our measurement of $\nu_{\rm max}$,
the frequency of the envelope peak, and column~4 gives the peak height (the
amplitude per radial mode, as plotted in Fig.~\ref{fig.amp.stars}).
Column~5 gives the peak amplitude of the $l=1$ modes, using the calibration
factor of $S_1/S_0 = 1.35$ (see Table~\ref{tab.spatial}), which we include
to emphasise that the strongest peaks will generally be those with $l=1$.

\subsection{The uncertainty in measuring oscillation amplitudes}  \label{sec.scatter}

Part of the uncertainty in each measured oscillation amplitude arises from
the precision with which we can fit and subtract the background.  This is a
measurement error and is given in columns 4 and 5 of Table~\ref{tab.amps}.

In addition, stars will presumably have variations with activity cycle,
analogous to the $\sim\pm5\%$ variation seen in the Sun, but for the moment
we do not have enough information to estimate those effects.  However, we
can estimate the scatter on the measured amplitude that is caused by the
finite lifetime of the modes.  This arises because a typical stellar
observing run does not last long enough to sample a large number of mode
lifetimes.

For the BiSON data in Fig.~\ref{fig.bison.amp.series}, the relative scatter
of the measured amplitude about the smoothed curve has an average value of
$\sigma_A/A = 6.2\pm0.2\%$.  This relative scatter $\sigma_A/A$ depends on
two parameters.  One is $T/\tau$, where~$T$ is the duration of the
observations (10\,d in this case) and~$\tau$ is the mode lifetime.  The
other parameter is~$N$, the number of 
low-degree oscillation modes that are excited within the smoothed envelope,
which is about 30 in the Sun.  We expect $\sigma_A/A$ to vary inversely
with both of these parameters:
\begin{equation}   \label{eq.scatterN}
  \frac{\sigma_A}{A} = k
  \left[N\left(1+\frac{T}{\tau}\right)\right]^{-0.5}. 
\end{equation}
We have verified that equation~(\ref{eq.scatterN}) holds, and that the
constant of proportionality is $k=0.75$, by using simulations with
different values of the parameters.  We simulated the stochastic nature of
the oscillations using the method described by \citet{DeRAK2006}.

Our decision to adopt a smoothing width of $4\Dnu$ (\S\ref{sec.amp}) means
that $N$ will be the same for all stars (provided there are not a large
number of additional mixed modes in the power spectrum).  Since we know the
mode lifetime of the Sun, we can calibrate equation~(\ref{eq.scatterN})
using the solar value of $\sigma_A/A = 6.2\pm0.2\%$ for the 10-d segments
of BiSON data.  The average mode lifetime in the Sun, measured over a solar
cycle from the linewidths in the range 2.8--3.4\,mHz, is $2.88\pm0.07$\,d
\citep{CEI97}.  Using this, we get
\begin{equation}  \label{eq.scatter}
  \frac{\sigma_A}{A} = (0.131 \pm 0.004)
  \left(1+\frac{T}{\tau}\right)^{-0.5}. 
\end{equation}
Equation~\ref{eq.scatter} can be used to estimate the uncertainty of
amplitude measurements that are made on other stars, provided~$\tau$ is
known \citep[see also][]{T+A94}.

The last column of Table~\ref{tab.amps} shows $\sigma_A/A$ for the
observations.  We have used published measurements of mode lifetimes for
\acena{} and~B \citep{KBB2005}, \nuind{} \citep{CKB2007} and \bhyi{}
\citep{BKA2007}.  There are no measurements available for the other stars
(\gser, \baql, \afor{} and \dpav) and so we adopted the solar value.

To obtain the total uncertainties for the amplitudes in
Table~\ref{tab.amps}, the uncertainty from the background subtraction
should be added in quadrature to $\sigma_A/A$ (although in practice, the
latter dominates).  Finally, we repeat that there is an additional source
of uncertainty in these amplitude measurements, namely any variations with
stellar activity cycle.

\section{Conclusions}

We have used daytime spectra of the blue sky, obtained with the stellar
spectrographs HARPS and UCLES, to measure oscillations in the Sun.  We
measured amplitudes by smoothing in power density and subtracting the
background, as previously described by \citet{KBB2005}.  In this paper we
propose two conventions in the use of this method.  Firstly, we suggest the
smoothing be done by convolution with a Gaussian having a FWHM (in power)
of~$4\Delta\nu$.  Secondly, we have chosen to report mean amplitudes for
the radial modes ($l=0$), but stress that the highest peaks in the
amplitude spectrum will generally correspond to $l=1$.

In the daytime sky observations reported here, the solar oscillation
amplitude was slightly lower than that measured simultaneously by BiSON, by
a factor of $1.07\pm0.04$.  This difference arises from two factors:
(i)~the stellar techniques measure a velocity that is dominated by neutral
iron lines, which are formed lower in the solar atmosphere than the
potassium line measured by BiSON; (ii)~the two methods have slightly
different spatial response functions.

A single-epoch measurement of the solar amplitude has a significant
uncertainty due to the stochastic nature of the oscillations.  We find the
mean solar amplitude measured by BiSON over one full solar cycle (for
radial modes) to be $20.0\pm0.1$\,\cms, implying a mean amplitude measured
using stellar techniques of $18.7 \pm 0.7$\,\cms.  The mean peak amplitude
for $l=1$ modes is $25.2\pm 0.9$\,\cms.  

By using simulations, we have estimated the scatter in the amplitude
measured from solar-like oscillations that arises from the stochastic
nature of the excitation and damping.  The result is given in
Equation~\ref{eq.scatter}, which matches the scatter that we find from
BiSON observations of the Sun.  This equation can be used to estimate the
uncertainty of amplitude measurements that are made other stars, provided
the mode lifetime is known.

Finally, Table~\ref{tab.amps} gives amplitudes for three stars (\gser,
\baql{} and \afor) that were measured from a series of short observations
with HARPS, together with revised amplitudes for five other stars for which
we have previously published results.  Now that the solar amplitude is
established, these measurements should be valuable tests for theoretical
models of solar-like oscillation amplitudes.

\acknowledgments

We thank Graham Verner for providing the time series of BiSON velocity
measurements, and Yvonne Elsworth and Dennis Stello for comments on this
paper.  We thank ESO and the AAO for supporting the daytime observations.
This work was supported financially by the Danish Natural Science Research
Council and the Australian Research Council.  We further acknowledge
support by NSF grant AST-9988087 (RPB), and by SUN Microsystems.  CK
acknowledges support from the Danish AsteroSeismology Centre and the
Instrument Center for Danish Astrophysics.

\begin{table*}
\small
\begin{center}
\caption{\label{tab.spatial} Spatial response functions}
\begin{tabular}{lccccc}
\tableline
\noalign{\smallskip}
\tableline
\noalign{\smallskip}
 & 
BiSON &
Stellar &
\multicolumn{3}{c}{Intensity} \\
 & 
velocities &
velocities &
402\,nm &
500\,nm &
862\,nm \\
\noalign{
\smallskip}
\tableline
\noalign{
\smallskip}
$S_1/S_0$\ldots & 1.37  & 1.35 &  1.26  &  1.25  &  1.20    \\
$S_2/S_0$\ldots & 1.08  & 1.02 &  0.81  &  0.75  &  0.67    \\
$S_3/S_0$\ldots & 0.58  & 0.47 &  0.25  &  0.18  &  0.10    \\
$S_4/S_0$\ldots & 0.22  & 0.09 &$-0.03$ &$-0.06$ &$-0.10$   \\
\noalign{									 
\smallskip}                      		        
\tableline									 
\noalign{									 
\smallskip}                      		        
$c$\ldots       & 4.43  & 4.09 &  3.31  &  3.16  &  2.91 \\
\noalign{
\smallskip}                             
\tableline                             
\end{tabular}
\end{center}
\end{table*}

\begin{table*}
\small
\begin{center}
\caption{\label{tab.amps} Amplitudes of velocity oscillations }
\begin{tabular}{lrlrrr}
\tableline
\noalign{\smallskip}
\tableline
\noalign{\smallskip}
 & 
$\Dnu$  & 
$\nu_{\rm max}$  & 
\multicolumn{2}{c}{$v_{\rm osc}$ (cm\,s$^{-1}$)} &
 \\
Star       &
(\muHz) &
(mHz) &
\multicolumn{1}{c}{$l=0$} &
\multicolumn{1}{c}{$l=1$} &
$\sigma_A/A$ \\
\noalign{
\smallskip}
\tableline
\noalign{
\smallskip}
Sun\ldots      & 134.8& 3.1  & $18.7\pm0.7$ & $25.2\pm0.9$ &  0.3\%\\
\gser\ldots    &  90  & 1.6  & $34.2\pm1.3$ & $46.2\pm1.8$ & 10\% \\
\afor\ldots    &  60  & 1.1  & $34.8\pm0.9$ & $47.0\pm1.2$ &  7\% \\
\baql\ldots    &  30  & 0.41 & $48.8\pm1.1$ & $65.9\pm1.5$ &  9\% \\   
\acena\ldots   & 106.2& 2.4  & $22.5\pm0.2$ & $30.4\pm0.2$ &  8\% \\
\acenb\ldots   & 161.4& 4.1  & $ 7.2\pm0.1$ & $ 9.7\pm0.2$ &  7\% \\
\bhyi\ldots    &  57.5& 1.0  & $41.9\pm0.3$ & $56.6\pm0.5$ &  6\% \\
\nuind\ldots   &  25.1& 0.32 & $64.7\pm0.9$ & $87.4\pm1.3$ & 10\% \\
\dpav\ldots    &  93  & 2.3  & $19.5\pm0.4$ & $26.3\pm0.5$ &  8\% \\
\noalign{
\smallskip}                             
\tableline                             
\end{tabular}
\end{center}
\end{table*}

\end{document}